\documentstyle[12pt]{article}

\begin{document}
\thispagestyle{empty}

\begin{center}
\LARGE \tt \bf{Spin polarised particles in G\"{o}del world}
\end{center}

\vspace{2.5cm}

\begin{center} {\large L.C. Garcia de Andrade\footnote{Departamento de
F\'{\i}sica Te\'{o}rica - Instituto de F\'{\i}sica - UERJ

Rua S\~{a}o Fco. Xavier 524, Rio de Janeiro, RJ

Maracan\~{a}, CEP:20550-003 , Brasil.E-mail:garcia@dft.if.uerj.br.}}
\end{center}

\vspace{2.0cm}

\begin{abstract}
The motion of classical test spinning particles in G\"{o}del universe in the realm of Einstein's General Relativity (GR) is investigated by making use of Killing conserved currents. We consider three distinct cases of motion of spinning particles polarized along the three distinct axes of the anisotropic metric. It is shown that in the case the spin is polarised along the $y-direction$ the minimum energy of the motion is attained for only for spinless particles while the other two directions the minimum energy is obtained for spinning particles. The continuos energy spectrum is also computed.
\end{abstract}

\newpage
\pagestyle{myheadings}
\markright{\underline{Spinning particles in G\"{o}del world}}
\section{Introduction}
\paragraph*{}

Recently  Figueiredo \cite{1} has discussed  the motion of quantum spinning test particles
in the case of  G\"{o}del space-time in  Einstein-Cartan gravity. The need of quantum spin  instead of using simply classical spinning particles stems from the fact that as shown by Hehl \cite{2} and Yasskin and Stoeger \cite{3} only quantum spin through their appearence in isolated spinning particles like electron or neutrinos or in extended bodies with a net polarisation, interacts with Cartan torsion. With this motivation in this letter we drop the requirement of spinning quantum particles and consider just classical spinning particles by making use of GR instead of Einstein-Cartan theory of gravity. Recently another example \cite{4} of this application has been done in the case of spinning particles around spinning string in EC gravity since in this case the exterior of the spinning string is nothing bur a Riemannian spacetime instead of the usual Riemann-Cartan spacetime. In this letter we investigate three cases of polarisation along three distinct axis of the G\"{o}del anisotropic universe. It is shown  that a classical spinning polarised particle along the y-axis possess a minimum energy just in the case all spin components vanish. Thus one may say that the minimum energy is attained only for spinless particles, while in the other two directions the spin components do not vanish, thus one may say that spinninig particles possess a minimum energy orbit. We also show that in the case of $p_{z}=0$ the enrgy is negative and the motion is bound. Previously Novello et al \cite{5} have investigated the confinment of spinless test particles and investigate the geodesic confinment of spinless particles in the context of G\"{o}del metric. In some sense the work presented here is a generalization of their work. It is also interesting to call the attention to the fact that one of the motivations for our study is that Maeda et al \cite{6} have investigate the production of chaos in the motion of spinning particles around the Schwarzschild and Kerr geometries in GR.   
\section{Killing conserved currents in G\"{o}del spacetime}
The equations of motion of the spinning particle in GR were investigated by Papapetrou \cite{7} and Dixon \cite{8}. They are given by the set of equations 
\begin{equation}
\frac{d{x}^{\mu}}{d{\tau}}=v^{\mu}
\label{1}
\end{equation}
\begin{equation}
\frac{Dp^{\mu}}{D{\tau}}= {R^{\mu}}_{{\nu}{\rho}{\sigma}}({\Gamma}) v^{\nu} S^{{\rho}{\sigma}}-{K^{\mu}}_{{\nu}{\sigma}}v^{\nu} p^{\sigma}
\label{2}
\end{equation}
\begin{equation}
\frac{DS^{{\mu}{\nu}}}{D{\tau}}=p^{\mu}v^{\nu}-p^{\nu}v^{\mu}
\label{3}
\end{equation}
Here ${\tau}$ is the proper time of the orbit , while $S^{{\mu}{\nu}}$ ,$v^{\mu}$ and $p^{\mu}$ are respectively the spin tensor, four-velocity and four-momentum of the spinning particle. The proper time is chosen in such a way that $v^{\mu}v_{\mu}=-1$. The absolute derivative $\frac{D}{D{\tau}}$ on the RHS of equations (\ref{2}) and (\ref{3}) is the Riemannian derivative with respect to the Christoffel connection. The connection ${\Gamma}$ in the Riemannian connection. Since $p^{\mu}$ is no longer parallel to the velocity $v^{\mu}$ we need a supplementary condition called Dixon condition on the center of mass and given by 
\begin{equation}
p_{\mu}S^{{\mu}{\nu}}=0
\label{4}
\end{equation}
Here the mass of the spinning particle m is given by
\begin{equation}
{m}^{2}= -p_{\mu}p^{\mu}
\label{5}
\end{equation}
As in GR chaos case the system considered here has several conserved quantities. In the Riemannian case ,the magnitude of the spin S is  given by
\begin{equation}
S^{2}:= \frac{1}{2}S_{{\mu}{\nu}}S^{{\mu}{\nu}}
\label{6}
\end{equation}
which are constants of motion. The Riemannian constants of motion reads 
\begin{equation}
C:= {\xi}^{\mu}p_{\mu}-\frac{1}{2}[{\xi}_{{\mu},{\nu}}]S^{{\mu}{\nu}}
\label{7}
\end{equation}
where the coma denotes partial derivative. In the next section we shall consider the angular-momentum constants from the Killing constants $C_{\xi}$ and the constraints equations. Here ${\xi}^{\mu}$ is the Killing vector obeying the well-known Killing equation 
\begin{equation}
{\xi}_{({\mu};{\nu})}=0
\label{8}
\end{equation}
where the semi-collon denotes the Riemannian covariant derivative. The G\"{o}del  line element would be written in retangular coordinates since this choice of coordinates proves to be easier to handle and to obtain the expressions of the spin components in terms of the linear momentum components. The G\"{o}del metric is given by 
\begin{equation}
ds^{2}= \frac{1}{2{\omega}^{2}}(-(dt+e^{x} dz)^{2}+dx^{2}+dy^{2}+\frac{1}{2}e^{2x}dz^{2})
\label{9}
\end{equation}
where ${\omega}^{2}$ is the rotation of the universe squared. Let us now consider the five-Killing vectors corresponding to the five isometries of the G\"{o}del spacetime \cite{9}
\begin{equation}
{\xi}^{\mu}_{1}=(1,0,0,0)
\label{10}
\end{equation}
\begin{equation}
{\xi}^{\mu}_{2}=(0,0,1,0)
\label{11}
\end{equation}
\begin{equation}
{\xi}^{\mu}_{3}=(0,0,0,1)
\label{12}
\end{equation}
\begin{equation}
{\xi}^{\mu}_{4}=(0,1,0,-z)
\label{13}
\end{equation}
\begin{equation}
{\xi}^{\mu}_{5}=(-2e^{-x},z,[e^{-2x}-\frac{1}{2}z^{2}],0)
\label{14}
\end{equation}
Substitution of these Killing vectors into the Dixon-Papapetrou general relativistic  conserved quantities, one obtains
\begin{equation}
E= -p_{t}+\frac{e^{x}}{4{\omega}^{2}}S^{zx}
\label{15}
\end{equation}
which is the total energy explicitly displaying the spin-orbit interaction. The other four conserved quantities are 
\begin{equation}
J= C_{3}= p_{z} -\frac{e^{x}}{4{\omega}^{2}}[S^{tx}-S^{xz}] 
\label{16}
\end{equation}
where J is the total angular momentum while the others are 
\begin{equation}
C_{5}=0= [-2p_{t}e^{-x}+p_{x}z+p_{y}(e^{-x}-\frac{1}{2}z^{2})+ \frac{e^{-x}}{2{\omega}^{2}}[S^{tx}-\frac{1}{2}S^{xz}e^{x}+ e^{-x}S^{yx}-e^{x}S^{yz}]]
\label{17}
\end{equation}
\begin{equation}
C_{4}= p_{x} - p_{z}
\label{18}
\end{equation}
\begin{equation}
C_{2}= p_{y}
\label{19}
\end{equation}
where $C_{5}$ is chosen to vanish for convenience.Where to perform this computation use has been made of the following nonvanishing components of the spin tensor $S^{{\mu}{\nu}}$ as
\begin{equation}
s= (S^{tx},S^{ty},S^{tz},S^{xy},S^{xz},S^{yz})
\label{20}
\end{equation}
where the spin tensor components satisfy the following property $S^{{\mu}{\nu}}=-S^{{\nu}{\mu}}$.
In the next section we shall examine three distinct cases corresponding to the case of motion of spin polarised motion along the three distinct axes $x$ ,$y$ and $z$ and analyse their properties.
\section{Spin polarised motion in G\"{o}del Universe}
In this section we shall be concerned with the motion of the spinning particles polarised along the three distinct directions. For example in the first case we shall consider the motion of the spinning particle confined at the plane $x-z$ where the spin of the spinning particle is polarized along the $y$-direction and
i) $p_{y}=0$. In this case we consider that the only non-vanishing components of spin tensor are $S^{xz}$, $S^{tx}$, $S^{ty}$ and $S^{tz}$ in principle. Substitution of this constraint into the center of mass expression (\ref{4}) yields 
\begin{equation}
S^{tx}p_{x}+S^{tz}p_{z}=0
\label{21}
\end{equation}
which is equivalent to
\begin{equation}
S^{tz}=- S^{xz}\frac{p_{x}}{p_{z}}
\label{22}
\end{equation}
the other components of the center of mass equation yields
\begin{equation}
S^{yt}=0
\label{23}
\end{equation}
\begin{equation}
S^{xz}= \frac{4{\omega}^{2}p_{t}[J-p_{z}]}{p_{t}+p_{z}}
\label{24}
\end{equation}
\begin{equation}
S^{tx}= \frac{4{\omega}^{2}p_{z}[J-p_{z}]}{p_{t}+p_{z}}
\label{25}
\end{equation}
Substitution of the last expression into the center of mass component equation yields 
\begin{equation}
S^{tz}=  \frac{-4{\omega}^{2}p_{x}[J-p_{z}]}{p_{t}+p_{z}}
\label{26}
\end{equation}
These expressions allows us to compute the the spin components in terms of the linear momenta. To achieve these results we have used another constraint which is $C_{5}=0$. Substitution of expression (\ref{24}) into (\ref{16}) yields 
\begin{equation}
E= -p_{t}+\frac{e^{x}p_{t}[J-p_{z}]}{p_{t}+p_{z}}
\label{27}
\end{equation}In the following subsection we shall compute the analogous example in the case the spin is polarised along the 
the $x-axis$ which means that the only spatial component of spin that survives is $S^{yz}$. Therefore this case is
ii)$p_{x}= 0$
Thus in this case we have that $S^{xz}$ and $S^{yx}$ vanish and this allows us to simplify the Killing conserved currents in the following manner. From the center of mass equations
\begin{equation}
S^{xt}p_{t}+S^{xy}p_{y}+S^{xz}p_{z}=0
\label{28}
\end{equation}
Placing the constraints of the vanishing of the other spatial components of the spin tensor we obtain
\begin{equation}
S^{xt}p_{t}=0
\label{29}
\end{equation}
which yields $S^{tx}=0$. Substitution of this constraint into the $C_{5}=0$ expression yields
\begin{equation}
J= C_{3}= p_{z} - \frac{e^{x}}{4{\omega}^{2}}S^{tx} 
\label{30}
\end{equation}
From this last expression one obtains the expression for $S^{tx}$
\begin{equation}
S^{tx}= 4[J - p_{z}] e^{x}{\omega}^{2} 
\label{31}
\end{equation}
But from expression $S^{tx}=0$ one obtains
\begin{equation}
J= p_{z}
\label{32}
\end{equation}
From the remaining center of mass components one obtains
\begin{equation}
S^{yt}=-\frac{p_{z}}{p_{t}}S^{yz}
\label{33}
\end{equation}
\begin{equation}
S^{zt}=\frac{p_{y}}{p_{t}}S^{yz}
\label{34}
\end{equation}
To obtain these expressions one needs the component $S^{yz}$ which in turn is given by the $C_{5}=0$ equation
\begin{equation}
-2p_{t}e^{-x}+p_{y}(e^{-x}-\frac{1}{2}z^{2})- \frac{1}{2{\omega}^{2}}S^{yz}=0
\label{35}
\end{equation}
This expression allows us to write the $S^{yz}$
\begin{equation}
S^{yz}=[-2p_{t}e^{-x}+p_{y}(e^{-x}-\frac{1}{2}z^{2})] 2{\omega}^{2}
\label{36}
\end{equation}
Note that also  since the motion is confined to the $x=0$ plane then the last expression reduces to
\begin{equation}
S^{yz}=[-2p_{t}+p_{y}(1-\frac{1}{2}z^{2})] 2{\omega}^{2}
\label{37}
\end{equation}
and at the $z=x=0$ or y-axis this expresion reduces to the simpler form
\begin{equation}
S^{yz}=[-2p_{t}+ p_{y}] 2{\omega}^{2}
\label{38}
\end{equation}

Note finally that since $S^{xz}=0$ this time the total energy is $E=-p_{t}$ which allows us to say that the minimum energy for the spinning particle is not trivial in the sense that the particles need not be spinless in this case. Let us finally consider the case where
iii) $p_{z}=0$.
In this case the only nonvanishing component of the spin is $S^{xy}$. By analogy with the previous case we can compute the component $S^{tx}$ from the expression $C_{3}=J$ as
\begin{equation}
S^{tx}= - e^{-x}4J{\omega}^{2} 
\label{39}
\end{equation}
Also in this case the energy is minimum for the spinning particle motion. It is intresting to pointed out that we can considered an important application of the ideas discuss here if we relax the hypothesis of polarisation and consider just the possibility that the particles are just confined in G\"{o}del geometry at the initial instant of time. This could lead us to investigate the possibility of chaos and the precession of spin of the particles yieldind chaos. This study may appear elsewhere. 
\section*{Acknowledgement}
I am deeply grateful to Professor Yu N. Obukhov, Prof. P.S.Letelier and Professor I.D. Soares for helpful discussions on the subject of this paper. Financial support from CNPq. and UERJ is gratefully acknowledged.

\end{document}